\documentclass[usenatbib,fleqn]{mn2e}
\usepackage{amsmath}
\usepackage{epsfig}

\newcommand{\apj}{ApJ}
\newcommand{\aj}{AJ}
\newcommand{\mnras}{MNRAS}
\newcommand{\aap}{A\&A}
\newcommand{\apjs}{ApJS}
\newcommand{\apjl}{ApJL}
\newcommand{\araa}{ARA\&A}
\newcommand{\nat}{Nat}

\title[Radio source / ICM interactions]{The long-term effect of radio sources on the ICM}
\author[J.F.~Basson \& P.~Alexander]
        {J.F.~Basson,\thanks{Email:j.basson@mrao.cam.ac.uk \newline\hspace*{29pt} p.alexander@mrao.cam.ac.uk}
        P.~Alexander$^{\star}$ \\
        Cavendish Astrophysics, Madingley Road, Cambridge CB3 0HE}

\begin{document}
\maketitle
\begin{abstract}
We have performed 3D hydrodynamical simulations of FR-II radio sources
in $\beta$-profile cooling-flow clusters.  The effects of cooling of
the cluster gas were incorporated into a modified version of the
\textsc{zeus-mp} code.  The simulations followed not only the active phase of
the radio source, but also the long term behaviour for up to 2 Gyr
after the jets of the radio source were switched off.  We find as
expected that the radio source has a significant effect on the cooling
flow while it is active, however we also
find that the effects of the radio source on the cluster are
long-lived.  A buoyancy driven convective flow is established as the
remnants of the radio source rise through the cluster dragging
material from the cluster core.  Although the central Mpc of the
cluster reverts to having a cooling flow, this asymmetric convective
flow is able to remove the cool gas accumulating at the cluster
core and indeed there is a net outflow persisting for timescales 
of about an order of magnitude longer than the time for which the
source is active or longer.  
The convective flow may also provide a mechanism to enhance the
metallicity of the cluster gas at large cluster radii.
\end{abstract}

\section{Introduction}
There has been significant recent interest in the interaction between
radio sources and clusters.  This has been driven by observations with
both the \textit{ROSAT} and \textit{CHANDRA} satellites which have
demonstrated clearly that the presence of a powerful FR-II radio
source very significantly affects the ICM/IGM in the vicinity of the
source.  For example, for the best studied cases of Cygnus A
\citep*{Carilli et al,Smith et al} and 3C84 in Perseus
(\citealt{Fabian et al2000}; \citealt{Fabian et al2002}), cavities in
the X-ray emitting gas coincident with the radio cocoon are clearly
observed and for Cygnus A the effects of a strong bow shock can be
clearly seen.  Recent reviews of the X-ray observational data on
radio-source/cluster interactions are given in \citet{McNamara2002}
and \citet{McNamara et al2000}.  This interaction between the radio
source and the ICM has led a number of authors to consider whether
radio sources could solve the so-called ``cooling-flow problem''
(e.g. \citealt*{BinneyTabor, McNamara et al2000, Reynolds et al,
Churazov et al, Quilis et al, Bohringer et al, RuszkowskiBegelman,
BrighentiMathews, RHB}, hereafter RHB), in which the cool gas expected
to accumulate at the cluster centre as a result of cooling is not
detected.

A number of authors have discussed the heating of the ICM by radio
sources in the context of the overall energy budget for the cluster
(e.g. \citealt{BinneyTabor, KaiserAlexander1999}; RHB;
\citealt{Alexander} and references therein).  \citet{Churazov et
al2002} discuss the energy balance between the cooling flow and the
AGN on the basis that there is some feedback between the cooling flow
(which results in matter accreting onto a supermassive black hole) and
the resulting mechanical power of the AGN. This could also be
responsible for the intermittency of the radio source as inferred in
some observations \citep{McNamara et al2000_1, Fabian et al2000,
OwenEilek}. The energy input from powerful FR-II radio sources is
certainly sufficient to have a significant effect on the cluster,
however the lifetimes of FR-II radio sources are of order 100 Myr
(e.g. \citealt{AlexanderLeahy}), they are therefore transient events
in the lifetime of a cooling flow \citep{Fabian}.  An important
question is therefore whether a radio source can have a long-term
effect on the cluster; for this reason it is necessary to consider the
evolution of the remnants of dead radio sources (i.e. after the jets
cease energy injection into the lobes).  During these latter stages of
evolution the dynamics of the remnants of the radio source will be
dominated by buoyancy forces \citep{GullNorthover}.  A number of
studies have considered the evolution of buoyant plasma bubbles within
a cluster.  For example \citet*{SarazinBaumOdea} and \citet{Churazov
et al} developed models for 2A 0335+096 and M87 respectively where the
observed radio and X-ray structures are modelled by buoyant bubbles of
radio-emitting plasma which drag colder material that had been
deposited by the cooling flow outwards from the cluster core.  These
bubbles are Rayleigh-Taylor unstable and therefore form a mushroom
cloud type structure \citep{BruggenKaiser}. \citet*{Saxton et al} have
modelled the northern middle radio lobe of Centaurus A as a buoyant
bubble, and simulations of buoyant gas in a cluster environment have
been presented in 3D by \citet{Bruggen et al} and in high resolution
2D by \cite{BruggenKaiser2002}. In all of these studies gas was
injected continuously near to the centre of the cluster with zero
velocity and in pressure balance with the surrounding gas and the
system allowed to evolve.  Recently RHB have used a more realistic
model which involved setting up the initial radio plasma by simulating
a jet; these conditions were then used as the starting point of a new
simulation in which the jet activity was turned off.

Although the basic physical processes have been established by these
studies, the details of the hydrodynamical evolution of a cooling flow
cluster containing a radio source needs to be fully determined if we
are to answer the question about the long-term effects of radio
sources.  In this and a future paper we extend these studies.  We
follow RHB in simulating a radio source evolving in a cluster so that
the initial conditions are as realistic as is possible, however we use
a fully 3D simulation and also include cooling of the ICM.  The radio
source evolves into a cluster environment with an established cooling
flow, it is then turned off after a simulation time corresponding to
approximately 50 Myr and we then follow the evolution of the system
for a further 2 Gyr.

\section{Computational Approach}\label{sec:comp}

The simulation strategy was designed to model a radio-source event in 
a cooling-flow cluster atmosphere. 
The simulations presented here differ from those of RHB
in two important respects.  Firstly they are fully three dimensional
and secondly we include cooling of the cluster gas within
an initial cooling flow atmosphere.
The \textsc{zeus-mp} code
\citep*{StoneNorman, StoneNormanII, Stone et al, Norman} 
with an additional cooling term was used, and the 
simulations were performed in spherical polar coordinates 
$(r,\theta,\phi)$
with a grid size of $192\times 64\times 64$.
The grid spacing
was linear in $r$ and $\phi$, and ratioed in $\theta$. For a ratioed
grid, the size of a grid spacing is a constant multiple of the adjacent
grid spacing. 
A grid structure that is more closely packed along the jet
axis (the $z$-axis in these simulations) can therefore be achieved (RHB).

The cluster atmosphere was modelled by assuming an initial isothermal
gas distribution that followed a $\beta$-profile $\rho_{\rm{ICM}} (r)
= \rho_0/\left[1 + (r/r_0)^2\right]^{3\beta/2}$, where $r_0$ is the
core radius.  The gravitational potential is dominated by the dark
matter and the form of the dark matter potential well was chosen so
that the gas was initially in hydrostatic equilibrium.  The
self-gravity of the gas was ignored resulting in a significant
reduction in computational time.

Cooling of the cluster gas was implemented by adding a cooling term to
the energy equation, and using a composite cooling curve for the ICM
taken from the literature \citep*{SutherlandDopita, Puy et al, Tegmark
et al} and assuming a metallicity of $\log\left(N_{Fe}/N_H\right) -
\log\left(N_{Fe}/N_H\right)_{\sun} = -0.5$.  The \textsc{zeus-3d} code
provides some support for cooling, and the routines were used as the
basis for our implementation in \textsc{zeus-mp} (which is based on
\textsc{zeus-3d}). In addition to replacing the original cooling
function with the improved version discussed above, the numerical
accuracy of the routine was improved. The energy equation including
cooling is an implicit differential equation which is solved in the
original \textsc{zeus} code using an iterative Newton-Raphson (NR)
technique. Our modification is to include an additional robust
bisection technique to obtain a better initial estimate for the
solution in those cases where the initial NR converged slowly -- a NR
iteration was then performed from this improved initial guess.

The modified code was
tested by comparing the results from a simulation with a constant
density medium with an analytical solution and by allowing the cooling
flow cluster to evolve without the presence of a radio source; the
results were in excellent agreement with results in the literature
\citep*{Sulkanen et al}.

The radio source event was simulated by injecting two light
anti-parallel conical jets as in RHB. The computational domain is
bounded by an inner and outer spherical boundary. At the base of the
jet (the inner spherical boundary), outflow conditions are employed
except when the jets enter the computational grid, and at the outer
boundary, inflow of gas is allowed.  The gas injected into the
computational domain representing the jet material is subjected to the
same cooling as the ICM -- this is unphysical since the jet material
should cool via synchrotron emission.  However the jet material is
very low density and its cooling time greatly exceeds the simulation
time; therefore the jet material in effect forms a gas with an
infinite cooling time.

The parameters characterising the simulations are: the ratio of the
jet density to the density of the ICM ($\rho_{\rm{jet}} /
\rho_{\rm{ICM}}$, hereafter $\rho_{\rm{rat}}$), the Mach number of the
jet with respect to the ICM ($M$), the steepness and core radius of
the cluster gas ($\beta$, $r_0$), the length of time for which the jet
is on ($t_{\rm{on}}$), and the mass inflow rate of the cluster in the
absence of a radio source ($\dot{M}$) - taken through the spherical
surface $r=300~\rm{kpc}$ at $1~\rm{Gyr}$. The density ratio and Mach
number are both defined at the base of the jet.  In a forthcoming
paper we present an analysis of this parameter space using 2D
simulations.  Here we discuss two 3D simulations with parameters which
represent typical model radio sources, one with a power ratio between
the jet and the cooling flow of 291 (run 1), a radio source in a weak
cooling flow, and another with a power
ratio of 73 (run 2), a radio source in a strong cooling flow. 
Both simulations have $\beta=0.5$,
$r_0=100~\rm{kpc}$, and $t_{\rm{on}}=59~\rm{Myr}$. The other
parameters are listed in Table \ref{tb:parms}.

\begin{table}
\caption{Simulation parameters.}
\label{tb:parms}
\begin{tabular}{ccccc}
\hline
Simulation & $\dot{M}~[\rm{M}_{\sun}\rm{yr}^{-1}]$ & $M$ & $n_{\rm{ICM}}~[\rm{m}^{-3}]$ & $P_{\rm{jet}}/P_{\rm{cf}}$\\
\hline
run 1 & 8.50 & 100 & $10^3$ & 291 \\
run 2 & 996 & 100 & $10^4$ & 73 \\
\hline
\end{tabular}
\end{table}

The jets were turned on at $t=353~\rm{Myr}$, after a cooling flow had
been established, and they were turned off at $t=412~\rm{Myr}$. 
The simulations ran until a simulation time of $t=2.4~\rm{Gyr}$.
All simulations were performed on a 16-node Beowulf cluster of
1.3-GHz AMD processors with 512 Mb of memory per node.

\section{Results}\label{sec:results}
In Figure 1 we show the time evolution for run 2, the results for run
1 are similar.  We shall therefore illustrate our discussion by
considering only run 2 and return to compare the results of run 1 and
run 2 when we examine the evolution of the cooling flow.  Shown in the
figure are the density, internal energy (pressure) and velocity
structure in the cluster gas and radio source at three different
epochs: 401 Myr while the radio source is still active; 518 Myr
approximately 100 Myr after the jets have been turned off; 1.89 Gyr an
epoch since the jets were switched off which is approximately 25 times
the lifetime of the radio source.

\subsection{Evolution of the radio remnant and ICM}

The evolution of the system while the jets are switched on follows 
very closely the results from earlier simulations (see for example
RHB).  In Figure 1 the cocoon is clearly visible as
the low density region; the cocoon expands supersonically and
is surrounded by a strong bow shock.  After the jets are switched off
the over-pressured cocoon continues to expand and is preceded still by a
bow shock.  The expanding remnants soon reach pressure balance, however
in this stage they are very light compared to the external gas and
rise buoyantly through the cluster.  This general behaviour is
similar to recent studies by various authors (\citealt{Churazov et al,
BruggenKaiser, Quilis et al, Saxton et al}; RHB).
Here however as in RHB the initial conditions 
correspond more realistically to those pertaining to a dead radio source
since we have followed the initial evolution when the jets are active
and as shown by \citet{Alexander} this active phase results in a
substantial amount of gas being dragged from the centre of the
cluster by the expanding source.  The long-term interaction between
the buoyantly rising remnants and the cluster is an interplay between
the remnants continuing to drag gas out of the cluster and inflow
driven both by the cluster gas infilling between the remnants and
continued cooling within the cluster.

\setcounter{figure}{1}
\begin{figure*}
\epsfig{file=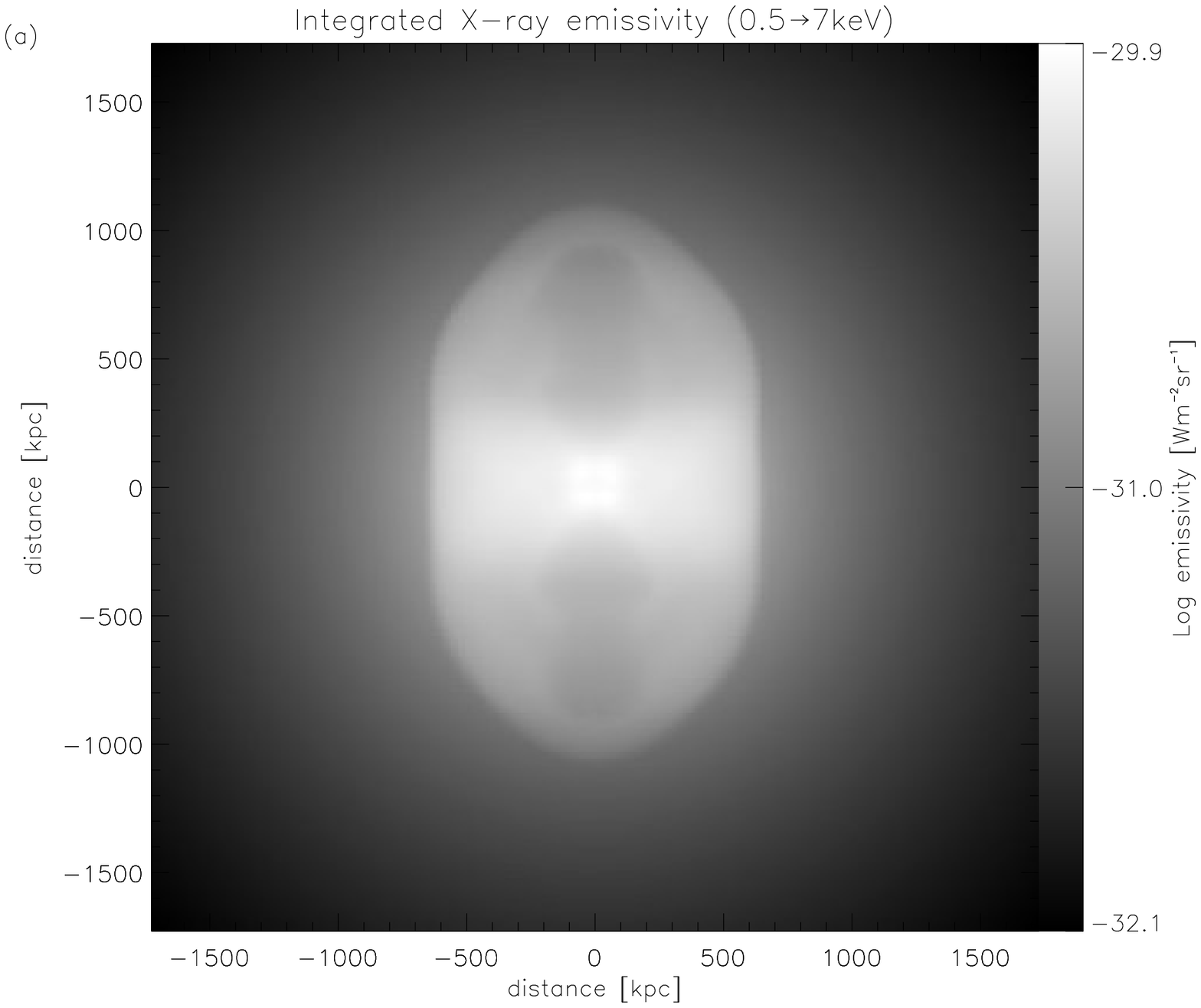,width=0.5\linewidth}\hfill
\epsfig{file=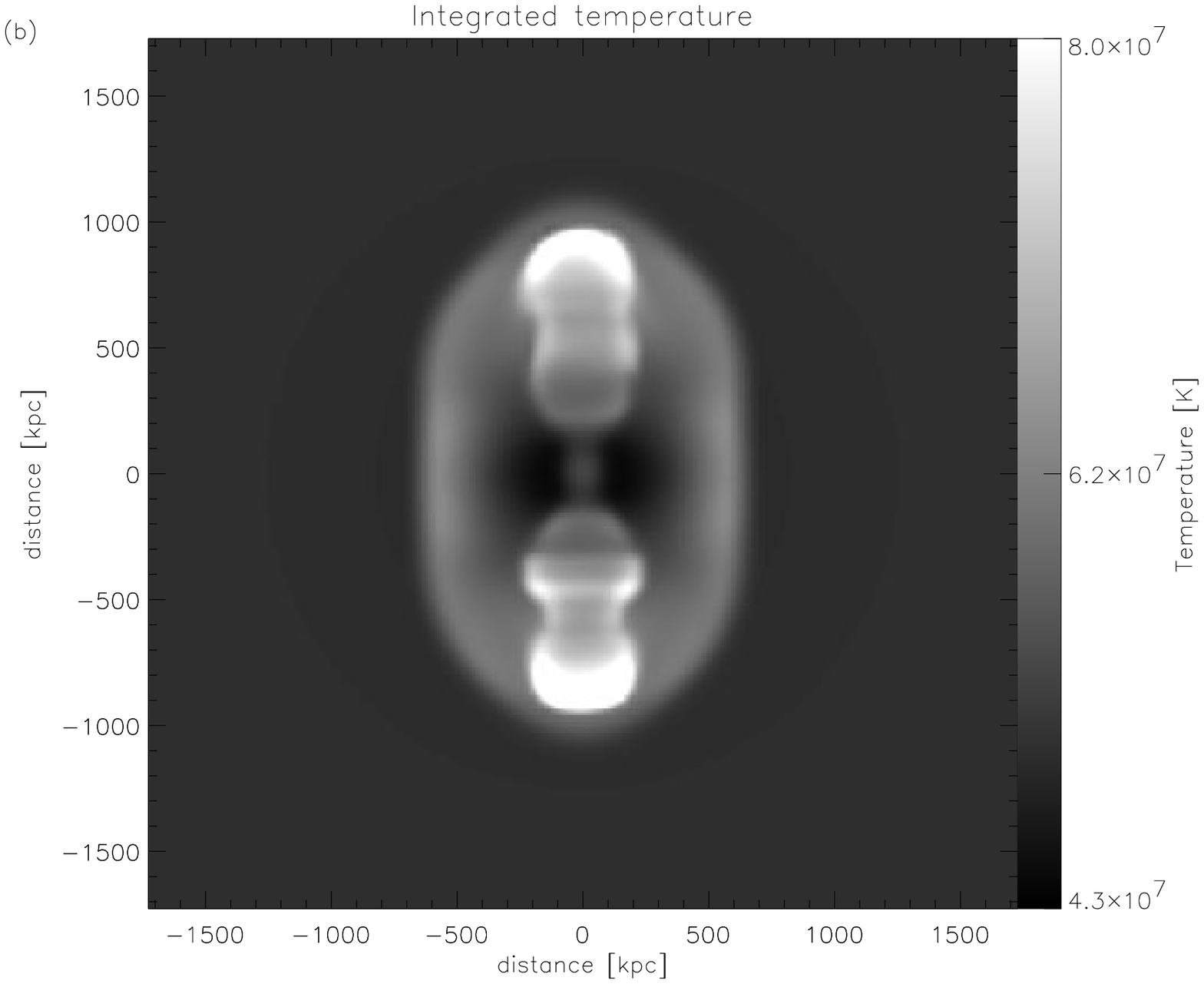,width=0.5\linewidth}
\epsfig{file=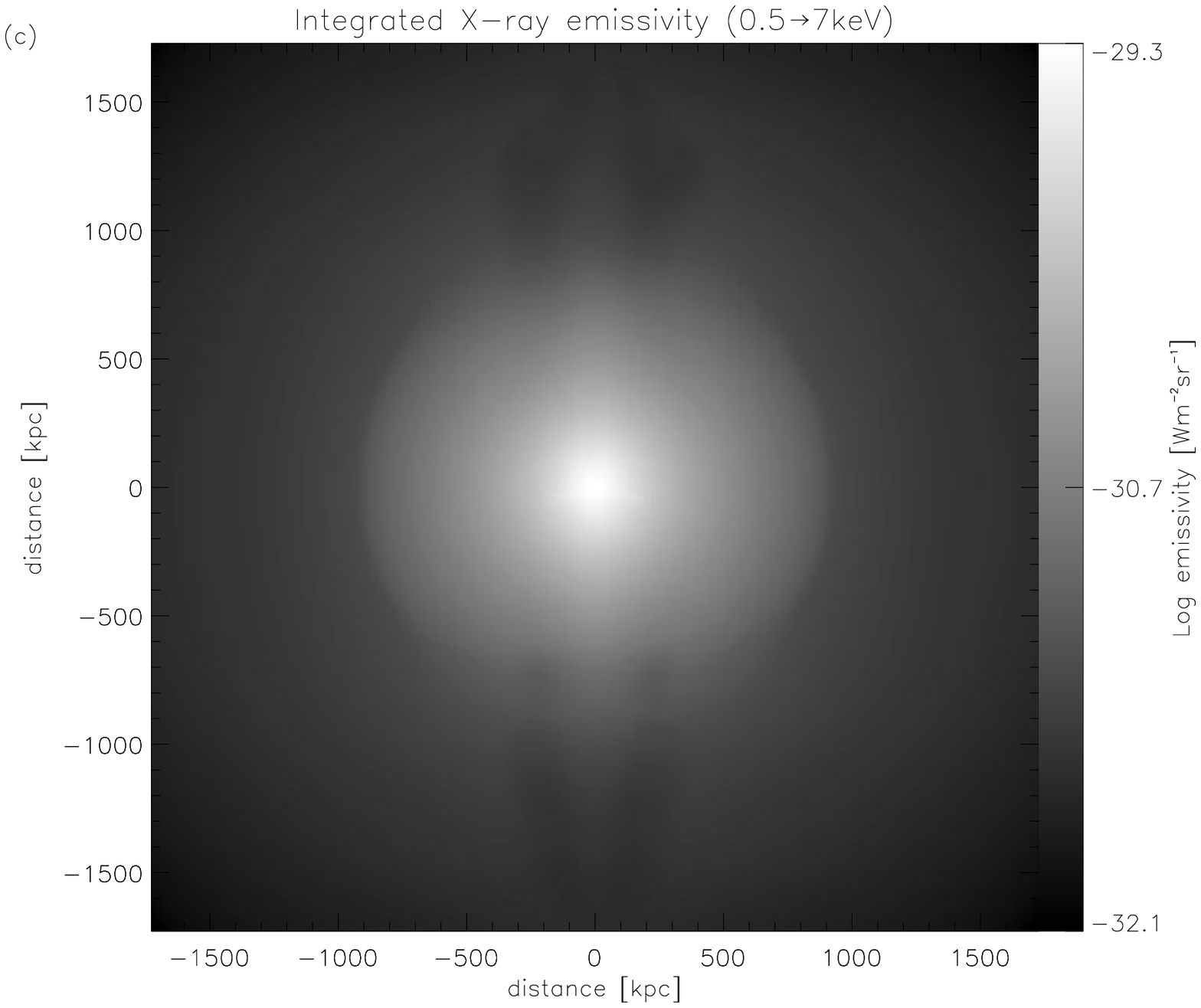,width=0.5\linewidth}\hfill
\epsfig{file=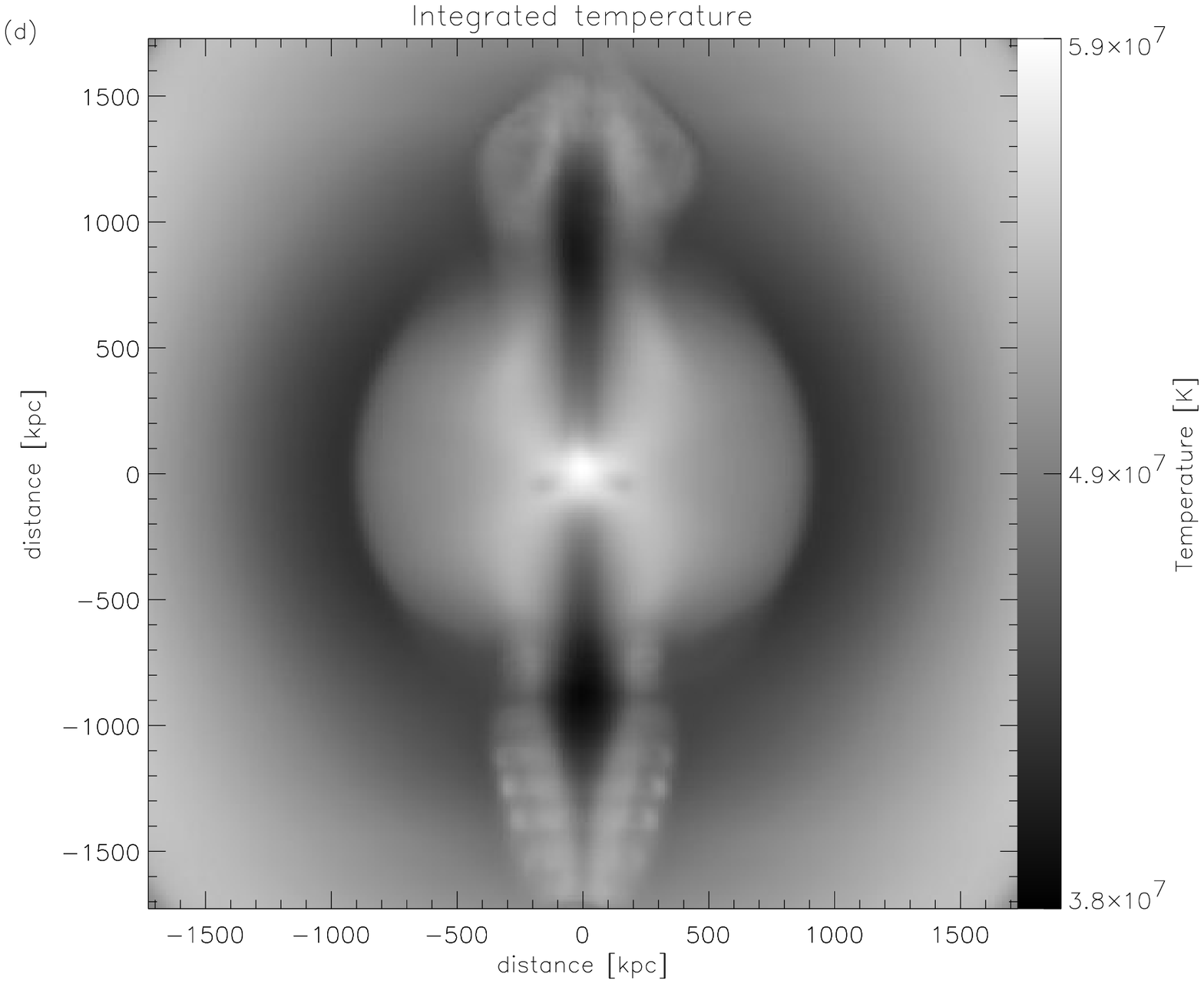,width=0.5\linewidth}
\caption{Integrated X-ray luminosity $\left(\iint \varepsilon_{\nu} d\nu dl;\, \varepsilon_{\nu} \propto n^2T^{-1/2}e^{-h\nu/kT}~\rm{Wm}^{-3}\rm{sr}^{-1}\rm{Hz}^{-1}\right)$ between 0.5 and 7 keV and emissivity weighted temperature $\left(\frac{\int T\varepsilon_{\nu}dl}{\int \varepsilon_{\nu}dl}\right)$ at 3 keV for run 2 at $707~\rm{Myr}$ ((a) and (b) respectively), and at $2.1~\rm{Gyr}$ ((c) and (d)).}
\end{figure*}

In Figure 2 we plot the integrated X-ray luminosity and temperature in the 
cluster gas for run 2 at two epochs after the jets have been switched off.

We shall now consider the evolution of the buoyant phase in a little
more detail.  Initially the buoyant remnants are surrounded by a weak
bow shock.  The simulated X-ray emission (Figure 2) is reminiscent of
the images recently obtained by \textit{CHANDRA} of radio sources in
cooling flow clusters (e.g. Perseus \citealt{Fabian et al2000};
\citealt{Fabian et al2002}; and A2052/3C317 \citealt{Rizza et al,
Blanton et al}).  In Figure 2b we show the emissivity-weighted
temperature integrated along the line of sight.  Immediately behind
the bow shock the temperature is higher than the surrounding gas,
however the temperature of the gas decreases systematically from
behind the shock towards the contact surface with the remnant radio
lobe.  This cooling is discussed in detail in \citet{Alexander} and
results from adiabatic cooling of the swept-up ICM as it is dragged
out of the cluster core. Simulations and analytical results presented
by \citet{BrighentiMathewsII} support the formation of cool gas by
expansion. The region occupied by the radio-emitting plasma here
appears saturated in the grey-scale.

At long times (after approximately 700 Myr) 
two main shocks are established within the ICM, 
which we shall refer to as the 
leading and trailing shocks (see Figure 1). 
The leading shock is again a weak bow shock
generated by the buoyantly rising remnants, while the trailing shock 
is produced by infalling ICM. The temperature structure
is now particularly interesting.  The ICM is heated slightly by the 
weak bow shock but then cools until the trailing shock; the infalling
gas is then heated by the trailing shock and within the cluster core 
there is little variation in temperature except in regions associated
with the radio source remnants themselves.  
The buoyant remnants undergo substantial mixing with the ICM as they
rise through the cluster. A shell of mixed ICM/radio plasma forms with 
a core of cooler ICM material (Figure 1 and Figure 2d) 
which is being dragged by the
remnants out of the centre of the cluster (see the velocity structure 
in Figure 1).  
The buoyant remnant is unstable to large scale instabilities and in
these simulations a large part of the remnant is shed
at approximately 2 Gyr.

The results presented here are in broad agreement with the
studies of buoyant radio plasma presented by a number of
authors \citep{Churazov et al, BruggenKaiser,
Quilis et al, Bohringer et al, Saxton et al}, however our results differ
in important details which are due to the initial conditions
we establish by following the evolution of a radio source
through its active phase.

\subsection{Mass flow and heating in the cluster}\label{sec:results.massflow}

In this Section we consider in detail the effect of the radio
source on the cooling flow.  
As pointed out by RHB approximately half the
energy input by the radio source goes directly into heating the
cluster gas.  This is a consequence of the self-similar
nature of the expansion.  The stored energy of the radio source
is just proportional to $p_c V_c$, while the work done by the
radio source on the ICM is $\int p_c dV_s$, where subscripts $c$ and
$s$ refer to the cocoon and shocked ICM respectively and we make use
of the fact that the shocked ICM is approximately at constant
pressure and equal to the pressure within the cocoon.
For the case of self-similar expansion $V_s \propto V_c$ and
$V_c$ and $p_c$ have a power-law dependence on time \citep{KaiserAlexander1997}.
Hence the stored energy in the cocoon and the energy input to the ICM
are proportional at all times and since the volume of the shocked gas
is comparable to that of the cocoon the constant of proportionality is
approximately unity.  Clearly therefore over the lifetime of the
source sufficient energy is input to significantly affect the cluster
as has been noted by many authors (e.g. 
\citealt{BinneyTabor,
Churazov et al,
BruggenKaiser,
Quilis et al};
RHB; \citealt{Alexander}).
The results presented here demonstrate that it is not only the heating
but also the redistribution of mass within the cluster which is
important in determining its evolution; this enables
the effects of a radio source event to affect a cooling flow on
timescales very long compared to the period over which the jet is active.

In order to quantify this effect we calculate the net radial mass flow
and cumulative flow
through a spherical surface of radius 300 kpc.
The results for runs 1 and 2 are shown in Figure 3.
\begin{figure*}
\epsfig{file=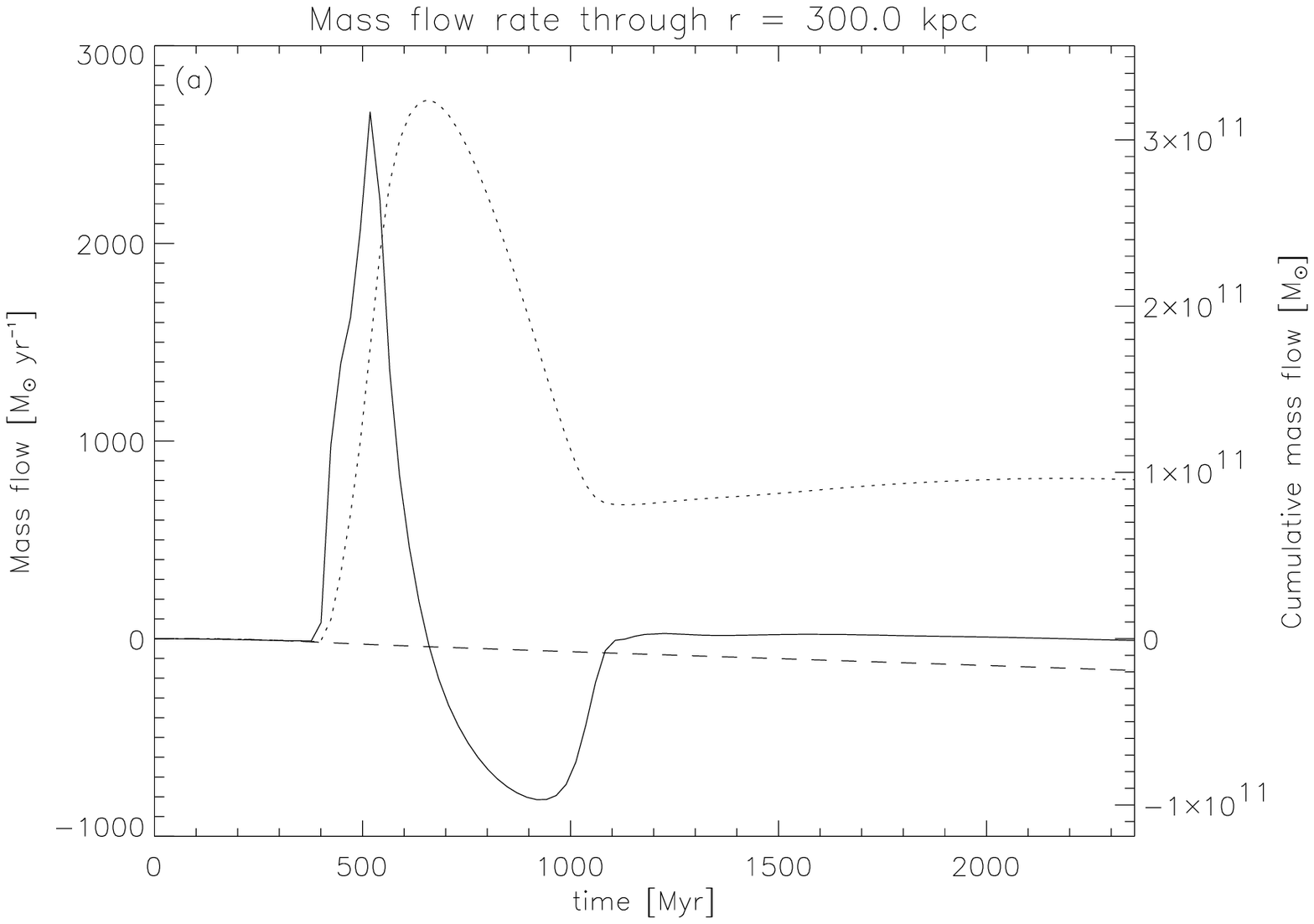,width=0.49\linewidth}
\epsfig{file=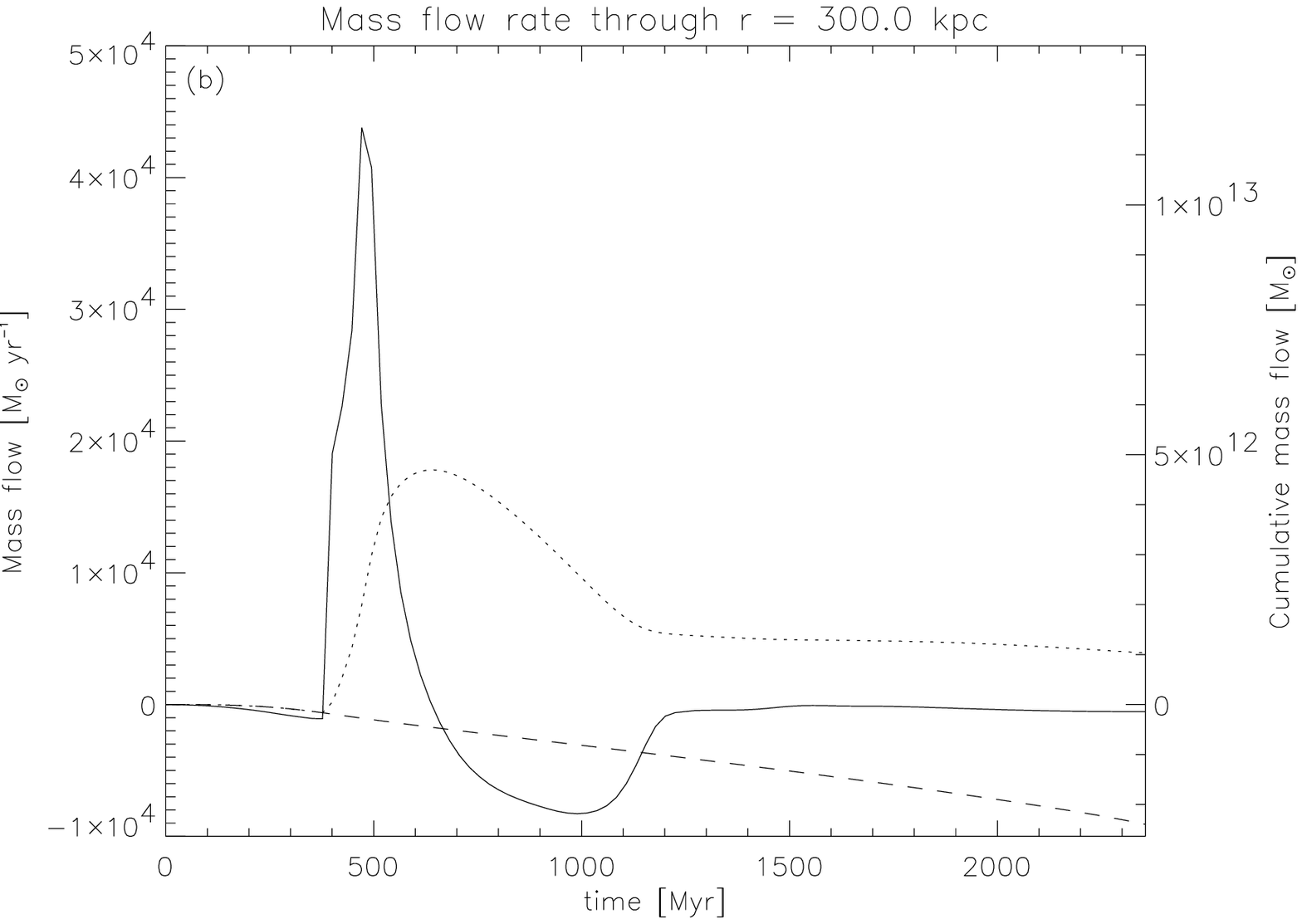,width=0.49\linewidth}
\caption{Mass flow $(\oint\rho\bf v\cdot dA)$ through the sphere $r=300~\rm{kpc}$ for run 1 (a) and run 2 (b). The solid line (values on left y-axis) is the instantaneous flow rate, and the dotted line (values on right y-axis) is the cumulative flow $(\int\negthickspace\oint\rho{\bf v\cdot dA}dt)$. The dashed line is the cumulative flow for the cluster in the absence of a radio source.}
\end{figure*}
The flow rate can be divided into 4 stages.  
Stage 1 is the cooling flow stage when the gas is flowing
steadily inwards (0 $\rightarrow 380~\rm{Myr}$). This is followed by
a sharp increase in gas flowing outwards as the shocked ICM and then
the cocoon material moves through the surface ($380~\rm{Myr}
\rightarrow 520~\rm{Myr}$). The third stage is a sharp increase in gas
flowing inwards, here the flow is dominated by the gas that lies
between the leading and trailing shocks which is moving towards the
centre of the cluster ($520~\rm{Myr} \rightarrow 940~\rm{Myr}$). 
The final stage depends on
the relative powers of the radio source and cooling flow: for
the lower power ratio of run 2 there is a return to cooling dominated inflow 
($> 1.5~\rm{Gyr}$),
although the inflow rate remains significantly less than for the same
cluster without a radio-source event; 
for the higher power ratio of run 1 there is always
a net outflow up until the end of the simulation (approximately 2.4 Gyr).
Similar behaviour exists for other surfaces with of course different
time evolution.

The long-term behaviour of run 2 is particularly interesting.  It is
clear from Figure 1 that even on these timescales the cluster is very
asymmetric.  Perpendicular to the original axis of the radio source
there is cooling driven inflow.  However, the buoyant remnants even at
times exceeding 1 Gyr are accompanied by a substantial wake of cluster
gas which is moving away from the cluster core.  For the strong radio
source this outward flow in the cluster gas exceeds the inwards
cooling flow.  Despite a cooling flow existing in most of the volume
of the inner Mpc of the cluster, the asymmetric flow (a remnant of the
earlier radio source event) ensures there is no accumulation of cold
gas at the cluster core.  In these simulations the cooler gas is
associated with the buoyant remnant at significant cluster radii.  The
cluster now contains a large-scale buoyancy driven convective flow.

\section{Discussion}\label{sec:discussion}

As has been noted by many authors 
(e.g. \citealt{Alexander} and references therein; RHB),
radio sources are certainly sufficiently energetic to provide
substantial heating to a cluster.  For example Perseus has an X-ray
luminosity of approximately $10^{37}$ W and powerful radio
sources have jet kinetic powers of order $10^{39}$ W, therefore if
the radio source lasts for a few $10^7$ yr the energy input from a
single powerful radio source is sufficient to provide all the
energy to power the X-ray emission for a substantial fraction of
a Hubble time.  However radio sources are short-lived events
and it is necessary to consider more carefully the balance between
heating rate and cooling rate.

Our results go some way to understanding the possible
role of radio sources in cooling-flow clusters.  Our simulations have
shown that the effect of a powerful radio source on its host cluster
is to modify the dynamics of the cluster gas on timescales 
very much longer than the lifetime of the active phase of the
radio source.  The buoyant remnants of the radio source drive a
large scale convective flow which is efficient at removing gas from
the cluster core on timescales comparable to a Hubble time.
By this mechanism the coolest gas will be moved to larger 
cluster radii.  Such a flow would also provide a mechanism for
metal enrichment of the cluster gas on large scales.

Observational support for these processes comes from the
\textit{CHANDRA} study of A133 \citep{Fujita et al}; this cooling-flow
cluster contains a ``radio relic'' which is connected to the host CD
galaxy by a tongue of cooler X-ray emitting gas.  The authors consider
the possibility that the X-ray structure is explained by uplifting by
the buoyant radio relic and demonstrate that this is feasible.  They
reject this possibility since the morphology of the cool X-ray
emission does not agree with the model predictions of \citet{Churazov
et al} or \citet{Bruggen et al}.  However, our simulations predict a
structure in the cooler X-ray gas (Figure 2 a\&c) which is remarkably
similar to the \textit{CHANDRA} results.  The structure in the X-ray
emitting gas depends strongly on the relative power of the radio
source and this will be explored in a forthcoming paper.

The simulations presented here although addressing the same problem as
RHB provide important new insight into the evolution of radio-source
containing clusters.  Firstly the buoyant remnants in the current
simulations show less mixing than in the simulations by RHB; the
result is that more of the available energy goes into large-scale
motion establishing the buoyancy-driven convective flow we have
described.  This is due in part to our simulations being 3D which
permit non-axisymmetric Kelvin Helmholtz instabilities to develop
which reduces the amount of mixing although we do see large-scale
instabilities.  Unfortunately the price we pay for going to 3D is to
loose resolution which will suppress high spatial frequency
instabilities which would be efficient at mixing.  However it is
possible that such modes may be suppressed in the real astrophysical
system by the magnetic fields which must be present within the radio
lobes.

The low grid resolution used here means that numerical diffusion could
be a problem. Here we argue that such numerical effects are small in
comparison with the overall behaviour observed in our
simulations. Firstly, we consider the relative importance of numerical
diffusion versus cooling by comparing the results of run 2 with a
simulation employing the same parameters, but without cooling (run
2a). For run 2a, the cumulative mass flow through a spherical surface
at $300~\rm{kpc}$ radius was $2.3\times 10^{12}~\text{M}_{\sun}$
compared to $1.0\times 10^{12}~\text{M}_{\sun}$ for run 2. This value
increases with time due to the convective effects of the rising
remnants for run 2a, compared with run 2 where this value is
decreasing with time as the cooling flow begins to dominate the
behaviour. We are therefore confident that the effects observed in the
simulation due to cooling are real, and are not being strongly
affected out by numerical diffusion of energy out of the computational
domain. A further check on the numerical accuracy is to compare
results at differing resolutions. In a forthcoming paper we will
present a 2D parameter space investigation, and also consider in
detail the effects of varying resolutions on the simulations. These
results show that the rate of energy loss through numerical diffusion
is essentially independent of the resolution used for resolutions in
the range $128\times64\rightarrow512\times512$ and overall energy
conservation in our simulations at all resolutions is as good as that
of RHB.

\section*{acknowledgements}
JFB is supported by the Commonwealth Scholarship Commission in the
United Kingdom.  We thank Helen Brimmer and David Titterington for
managing the Beowulf cluster.  We also thank Christian Kaiser and
Malcolm Longair for helpful discussions, and the referee for helpful
comments.

\clearpage

\newlength{\figheight}
\setlength{\figheight}{155pt}
\setcounter{figure}{0}
\begin{figure*}
\vspace*{20pt}
\epsfig{file=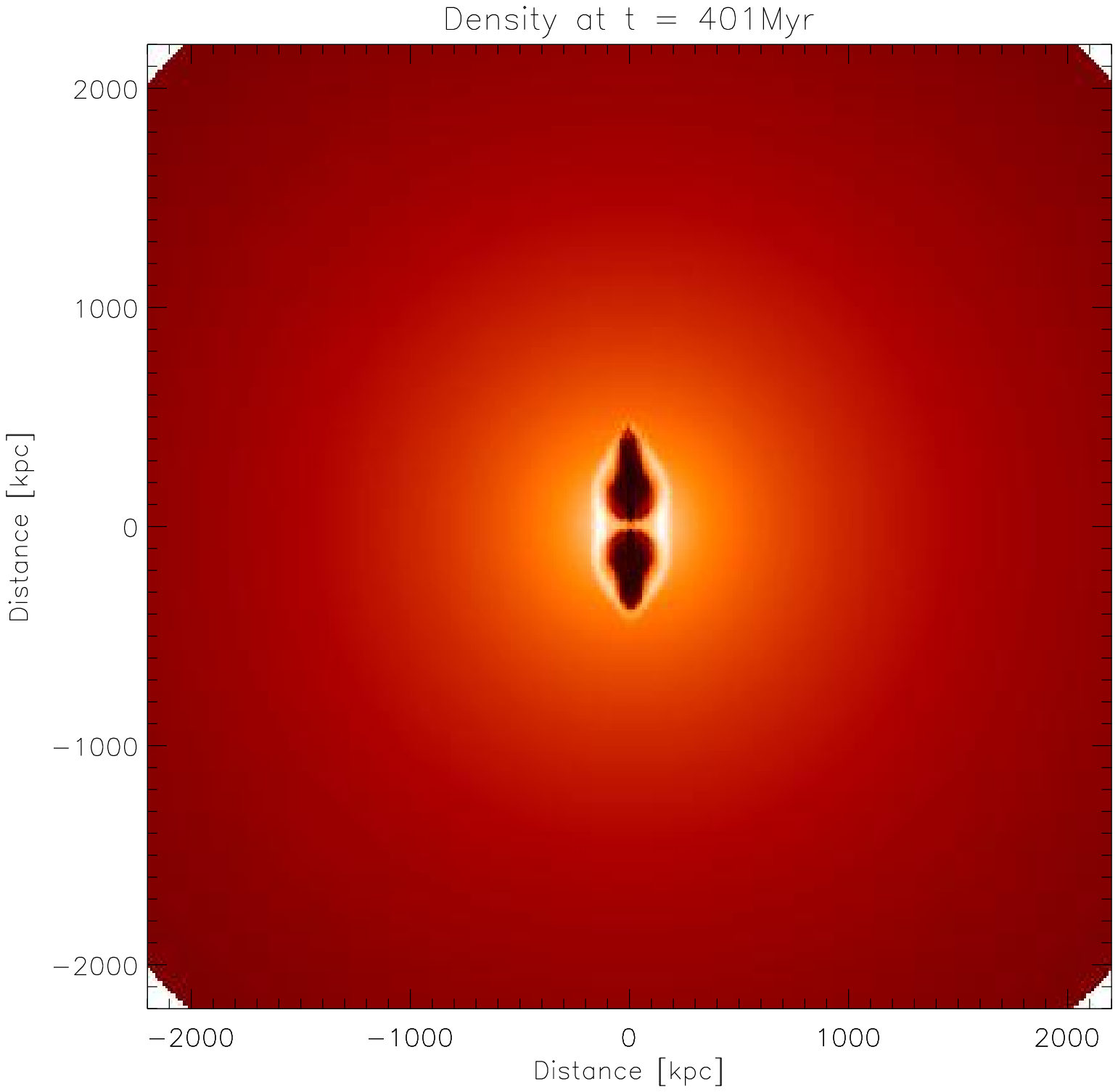,height=\figheight}\hfill
\epsfig{file=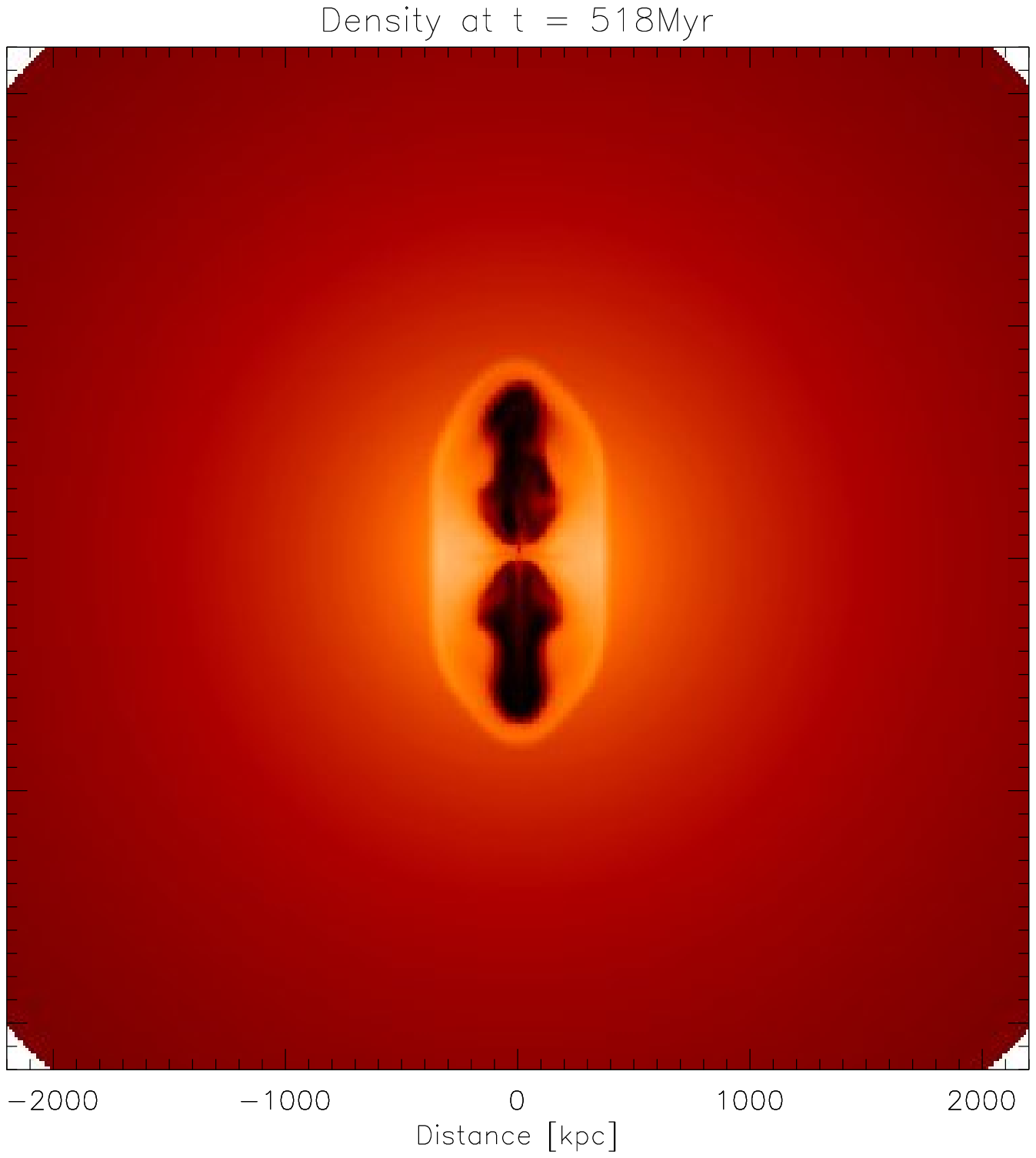,height=\figheight}\hfill
\epsfig{file=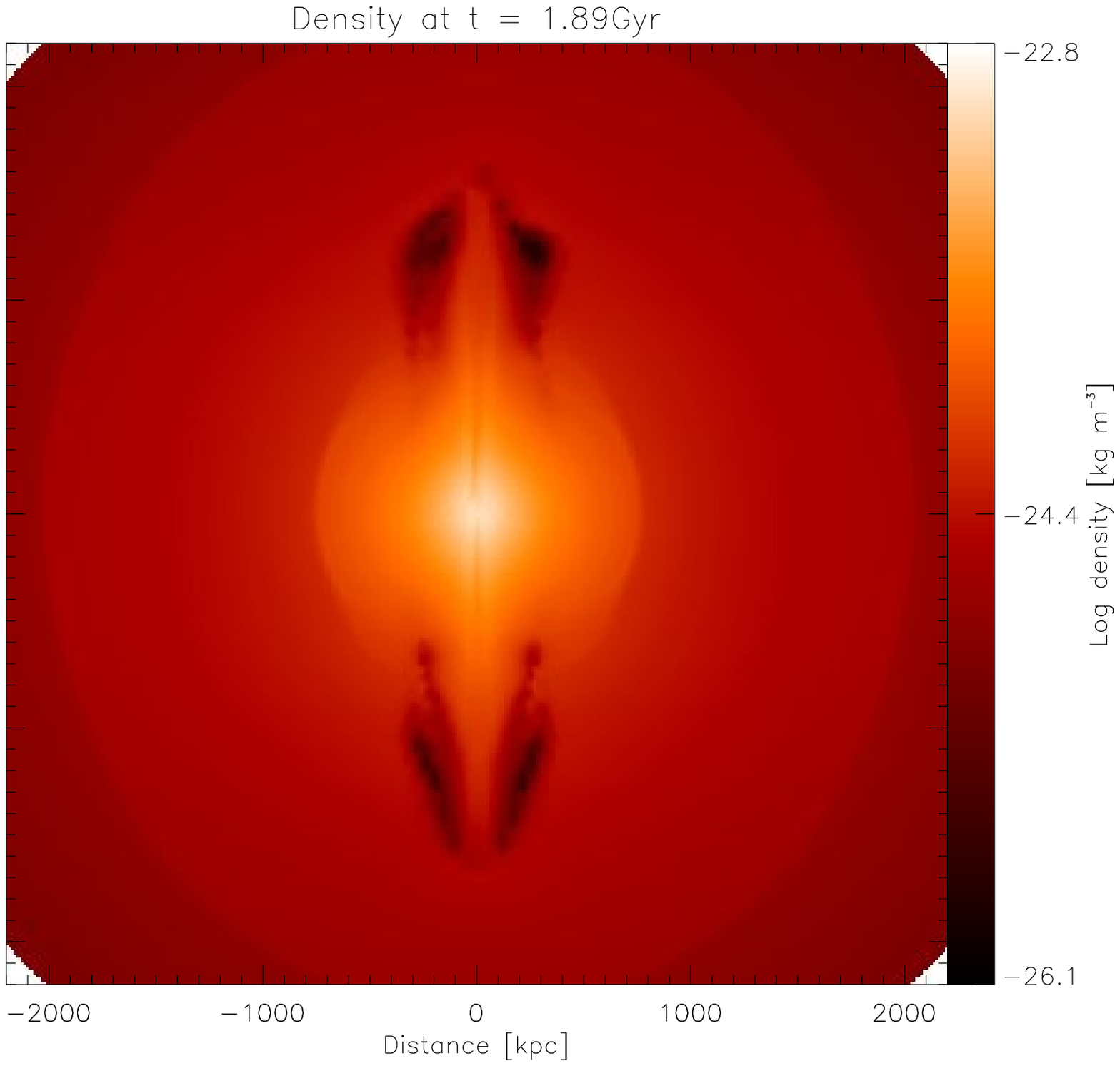,height=\figheight}

\vspace*{20pt}
\epsfig{file=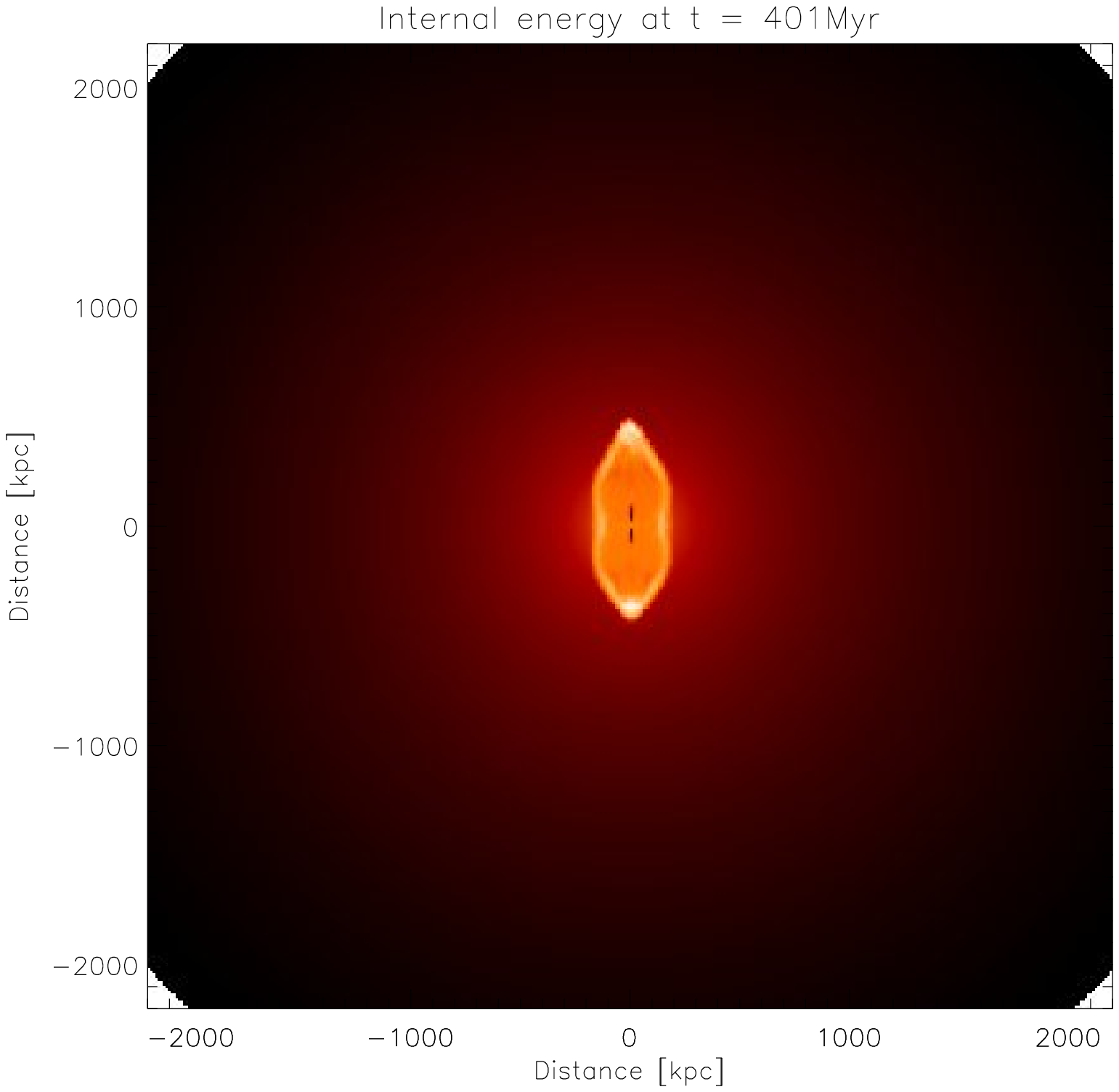,height=\figheight}\hfill
\epsfig{file=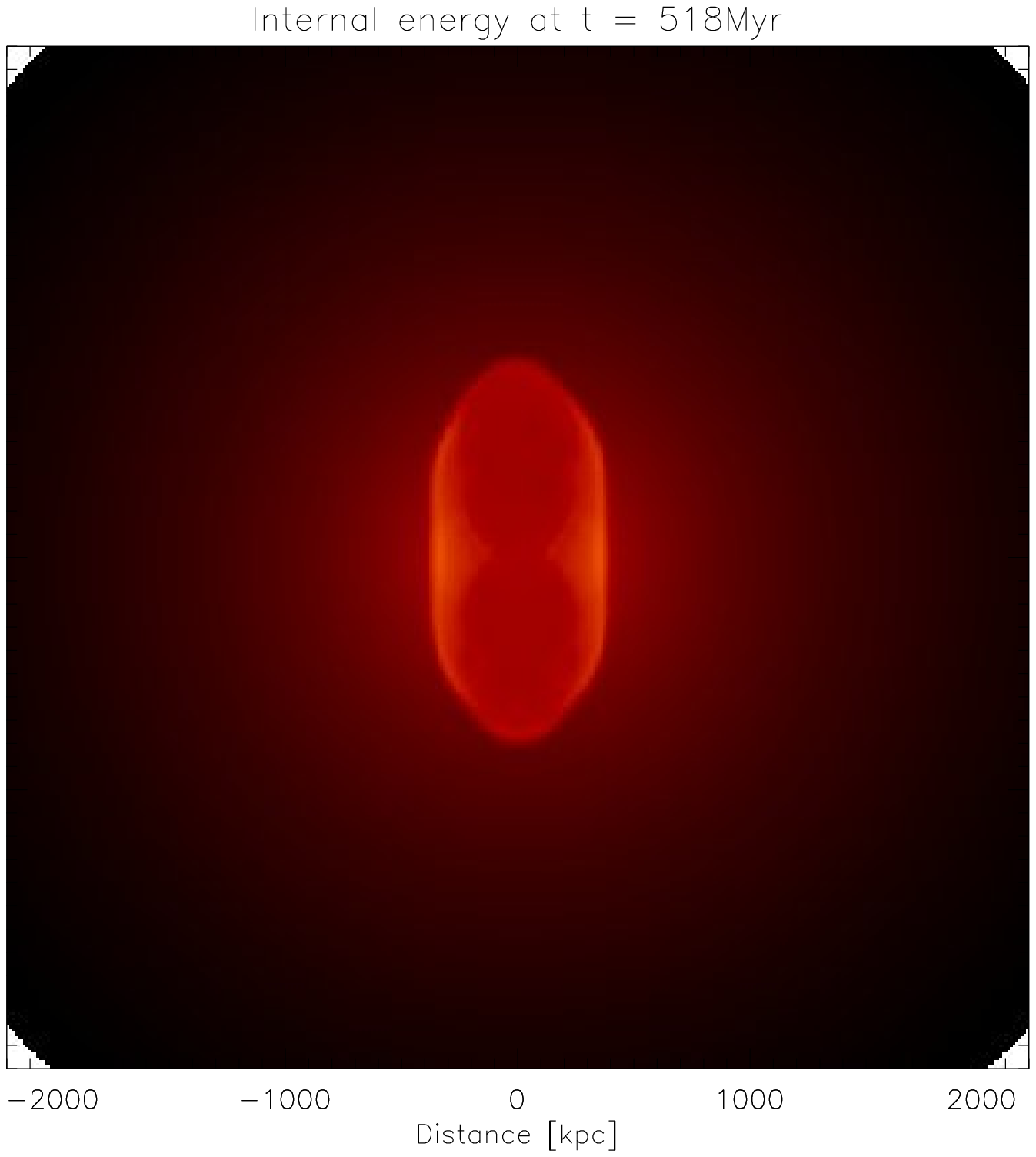,height=\figheight}\hfill
\epsfig{file=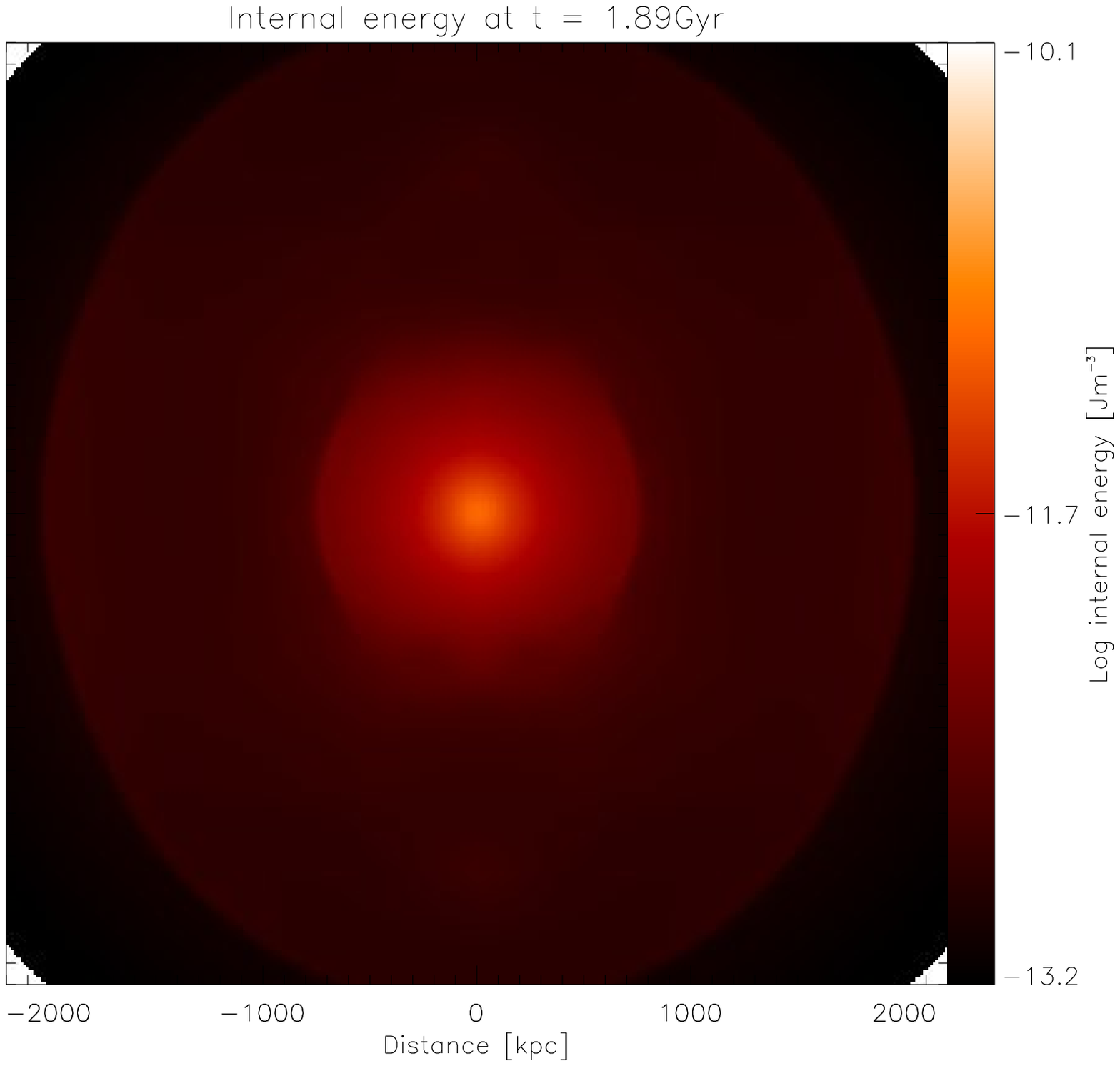,height=\figheight}

\vspace*{20pt}
\epsfig{file=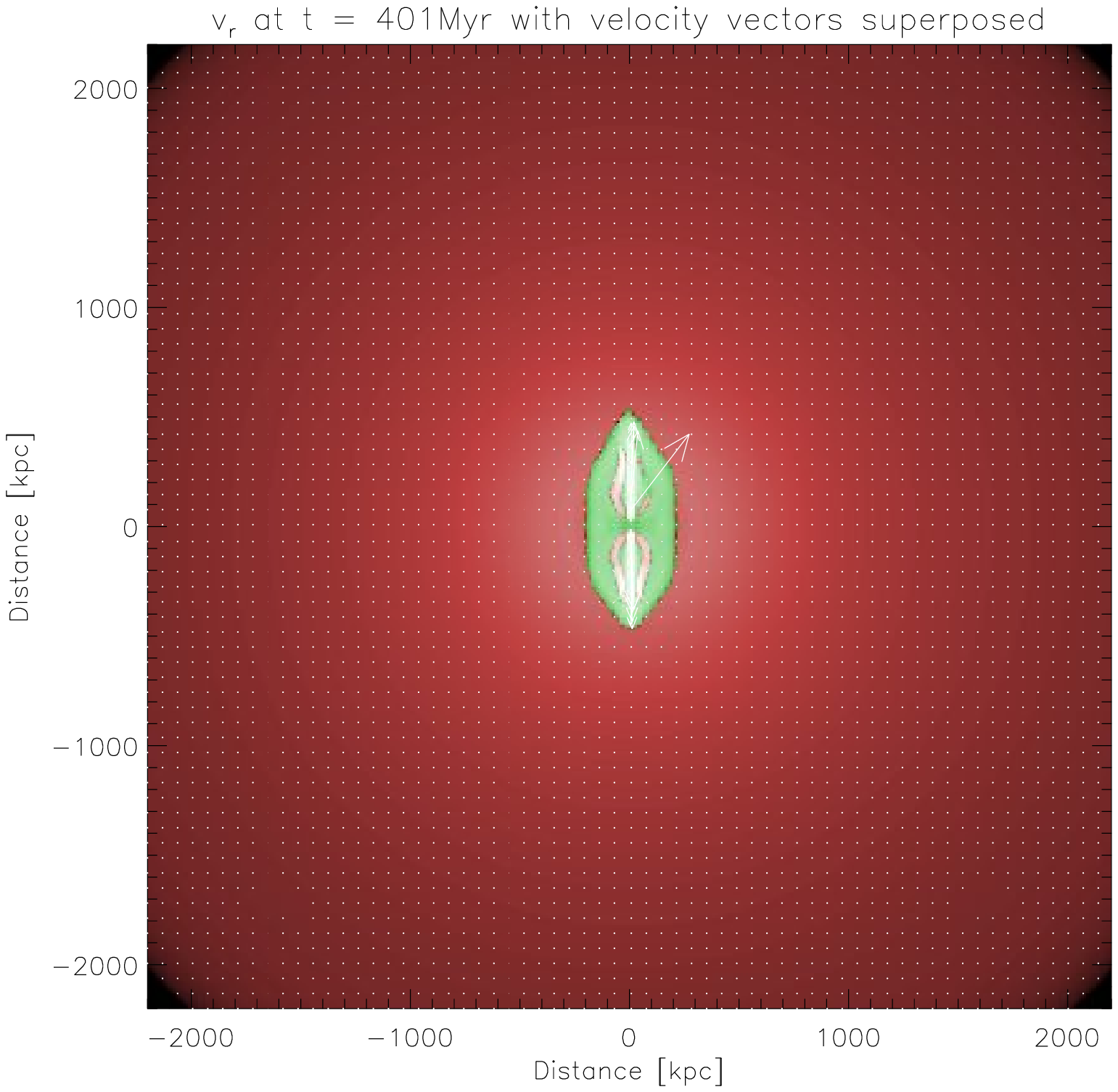,height=\figheight}\hfill
\epsfig{file=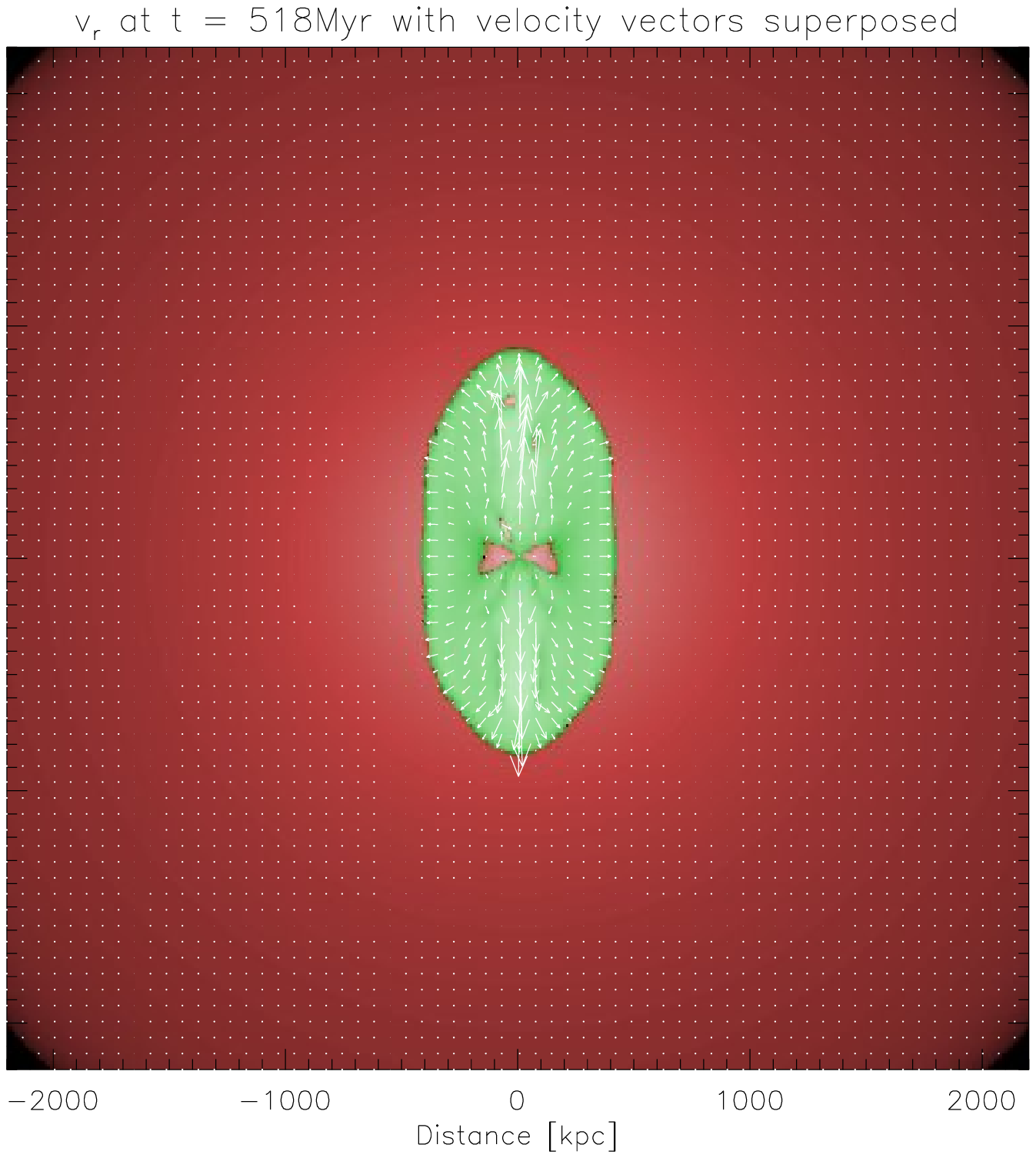,height=\figheight}\hfill
\epsfig{file=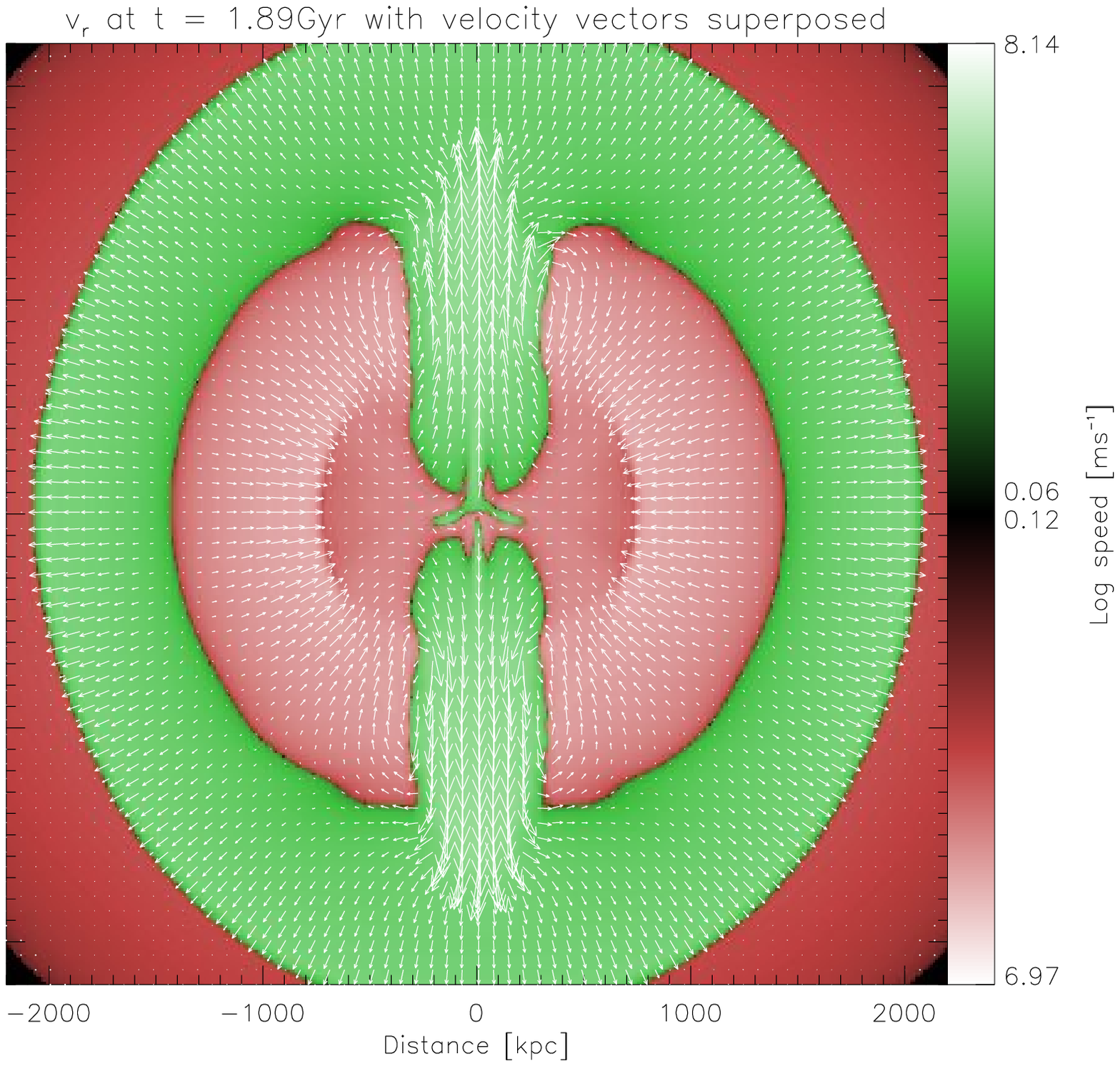,height=\figheight}
\caption{Density, internal energy, and velocity plots (top, middle and bottom rows respectively) at 401 Myr, 518 Myr and 1.89Gyr (left, middle and right columns respectively). The image in the velocity plots is the $\hat{r}$ component of the velocity, with red indicating negative velocity and green positive. The velocity vectors are overlaid.}
\end{figure*}

\end{document}